\documentclass{ws-procs975x65}

\def\beq{\begin{equation}}
\def\eeq{\end{equation}}

\begin{document}

\title{Fast Radio Bursts: Searches, Sensitivities \& Implications}
\author{Evan F. Keane$^*$ and the SUPERB Collaboration}

\address{SKA Organisation, Jodrell Bank Observatory,\\
Macclesfield, Cheshire SK11 9DL, United Kingdom\\
$^*$E-mail: Evan.Keane@skatelescope.org\\
www.skatelescope.org}

\begin{abstract}
Fast radio bursts (FRBs) are millisecond-duration transient signals
discovered over the past decade. Here we describe the scientific
usefulness of FRBs, consider ongoing work at the Parkes telescope, and
examine some search sensitivity and completeness considerations
relevant. We also look ahead to the results from ongoing and future
planned studies in the field.
\end{abstract}


\bodymatter

\section{Scientific Motivation}
Fast radio bursts (FRBs) have a story which has been told and retold
many times over the past few years. This will be summarized by others
in these proceedings\cite{matthew,emily}. Here it is sufficient to say
that FRBs are rare millisecond-duration radio bursts detected by high
time resolution sky surveys in the radio, and that they have sparked
excitement and controversy since they were first discovered in
2007. As is the case with many scientific endeavors there are perhaps
two questions that matter the most: (i) what are they? and (ii) what
are they good for?

There are those that have argued that FRBs are not astrophysical in
nature, and there have been suggestions that they are some form of
devilish radio frequency interference
\cite{bbe+11,kbb+12,bagchi,swiss}, although this line of inference
seems to have ceased with the successful identification of the
so-called `peryton' RFI signals\cite{pkb+15}. There are those who have
argued for\cite{lsm+14,mls+15} and against\cite{tuntsov14,dennison14}
a Galactic interpretation. There are those that have argued in favor
of an extragalactic interpretation
\cite{lbm+07,tsb+13,pp13,lg14,mz14,katz14,katz15,mj15,dgbb15,cw16,caleb16}.
Some have, over time, argued for all of the above. In getting to the
bottom of what the FRB progenitors are we should draw a comparison
with the quest to understand $\gamma$-ray bursts: then, as now, the
solution was to localize the sky coordinates of the signals to much
higher precision than had been previously done\cite{mdk+97}. Only then
can a meaningful comparison be made with the predictions of the by-now
very long list of proposed FRB models
\cite{zhang14,fr14,lyubarsky14,rl14,kon+14,kon15,brv14,iwazaki15,soko15,fo15,gh15,chiara15}.

As well as the challenge of identifying the progenitor, or
progenitors, for a new class of object, the main reason FRBs are a hot
topic is due to their potential utility as cosmic probes. If a large
number (hundreds to thousands) of FRBs can be detected, and be
localised such that their redshifts can be measured, then this
information can be used to (amongst other things) find out if the
``missing baryons'' are residing in the intergalactic medium, probe
intergalactic magnetic fields and measure the dark energy equation of
state parameter\cite{mkg+15}. These uses are independent of whether or
not FRBs are standard (or standardisable) candles; if they are then
there are further obvious applications\cite{rfc+98,pag+99}.

\section{Ongoing FRB searches and prospects}

The Southern hemisphere component of the High Time Resolution Universe
(HTRU) survey at Parkes\cite{kjs+10} ran from 2008 to January 2014. It
was very successful in its main objective of finding high-dispersion
measure millisecond pulsars at low Galactic
latitudes\cite{bates+11,burgay+13,bates+15}, also making several
noteworthy serendipitous discoveries such as identifying a magnetar in
the radio\cite{lbb+10} and the so-called ``diamond
planet''\cite{bbb+11}. Additionally HTRU South made significant
strides in searches for ``Lorimer bursts''\cite{lbm+07}, as FRBs were
then known, identifying at first four new examples\cite{tsb+13} which
markedly increased interest in the field. The Northern hemisphere
component of the HTRU survey began 2 years after its Southern
counterpart and is still ongoing. It too has discovered
pulsars\cite{bck+13} and continues to do so, but as yet has not
identified any FRBs. HTRU North has less time on sky by a factor of
$\sim 3$, and a narrower field-of-view by a factor of $\sim 4.5$;
however the gain is about twice that of the Southern survey. This
implies -- in the simplest estimate where we assume that the number of
sources $N$ scales with the flux density $S$ as $N \propto S^{-3/2}$
-- that the FRB expectation of HTRU North is $\sim 2$, in comparison
to HTRU South's final yield of $9$ FRBs\cite{champion15}.

Fueled by the results at Parkes, FRB search programs have sprung up at
other telescopes and in other observing bands; these have ranged from
dedicated projects to simply searching extant archival data. The first
FRB not discovered at Parkes, resulted from the re-analysis of
observations at 1.4 GHz taken at Arecibo\cite{sch+14}. A fast imaging
campaign searching for FRBs with 5-millisecond time resolution at 1.4
GHz has also been performed at VLA\cite{law15}, with no new detection,
likely due to as yet insufficient time spent on sky. Additional
experiments have then been planned, e.g. the new Apertif
system\cite{leeuwen14} --- based on phased array feed (PAF) receivers,
now at the commissioning stage at WRST --- offers a wealth of
potential FRB discoveries in the Northern sky at 1.4 GHz.

The FRB\cite{masui15} detected at the Green Bank radio telescope is
particularly interesting since it the first outside of the $\sim
1.4$-GHz band, showing that FRBs can be detected at $\sim
800$~MHz. This region of the radio spectrum, is where the Molonglo
Synthesis Telescope will survey the sky for FRBs. Molonglo is
currently being refurbished for high time resolution studies of the
sky\cite{caleb16a}, and when complete will perform continuous FRB
searches with an unrivaled product of time-on-sky and
field-of-view. If Molonglo's sensitivity is sufficient (which depends
on FRB spectra and the final system performance) then it could have an
FRB discovery rate twice that of
Parkes\cite{caleb16a}. CHIME\cite{bandura14}, a telescope under
construction in Canada, will have similar specifications to Molonglo
and operate at the intermediate frequency range
$400-800$~MHz. Finally, searches at much lower frequencies are ongoing
at LOFAR\cite{coenen14} and MWA\cite{tingay13}. However, as the shape
of the FRB spectra is still unclear, the prospects for detection at
frequencies below $\sim 800$~MHz remain uncertain as yet.

All in all, and except for Parkes, the various other endeavors are
either yet to be completed, or are suffering from different
combinations of insufficient time-on-sky, field-of-view and
sensitivity. Thus, as of early 2016, Parkes remains the premiere FRB
discovery instrument on the planet. At present the SUrvey for Pulsars
and Extragalactic Radio Bursts
(SUPERB\footnote{\texttt{https://sites.google.com/site/publicsuperb/}})
is underway at Parkes. This project arose at the end of 2013, when it
was clear that many improvements -- with respect to the aforementioned
HTRU survey -- could have been applied to enable a more optimized FRB
survey. By that stage GPU hardware had progressed and, crucially,
software to fully exploit it had been
developed\footnote{\texttt{http://sourceforge.net/projects/heimdall-astro/}}$^{,}$\footnote{\texttt{https://github.com/ewanbarr/peasoup}}.
It was thus feasible to perform FRB (and pulsar) searches over a wide
range of dispersion measure (and orbital accelerations). In
particular, SUPERB was devised with the following goals in mind:

\begin{itemize}
  \item Find more FRBs.
  \item Find out more information per FRB than ever before.
  \item Perform real time pulsar and transient searches.
\end{itemize}

Experience from past experiments had shown that a major obstacle for
constraining the nature of the FRBs was the long lag between the
occurrence of the radio burst and the starting of any campaign aimed
to identify an FRB counterpart in other electro-magnetic bands.

SUPERB was then designed in order to maximally reduce the dead time
between an event and its notification to the observers in the
community of the follow-up teams. In particular, a large-scale
multi-wavelength collaborative effort has been constructed in parallel
with the set-up of the survey. The advantages of a prompt reaction and
a rapid follow-up of the most interesting events became immediately
evident, when the first `peryton' signal occurred during a SUPERB
observation.  In fact, a quick and secure identification of the origin
of these RFI signals was finally possible. Similarly, there may be
cause for optimism that at least some FRBs identified with SUPERB may
be localized on the sky, and thus some of the numerous scientific
applications of FRBs may already be realizable. The earliest results
from the SUPERB survey are due to be published in 2016.

\section{Search Sensitivities \& Completeness}

At the time of preparing this contribution (early February 2016) there
were $N_{\mathrm{FRB}}=16$ FRBs in the literature\cite{frbcat}, covering the time period
2001 to 2014 inclusive. Although there are no FRBs yet known to have
been detected in 2002 to 2008 inclusive this mostly reflects the fact
that no large-scale wide-field pulsar/transient surveys were being
undertaken at Parkes during that time. The FRB Catalogue (FRBCAT) has
just been launched\footnote{\texttt{http://astronomy.swin.edu.au/pulsar/frbcat/}},
and should aid future work on population characteristics as the
numbers of FRBs grow. Indeed, one can already perform a number of important
studies with the information stored in FRBCAT.

For example one can investigate the reported Galactic latitude
dependence of FRBs\cite{psj+14,pkb+15}. The observations seem to
suggest a higher observed rate at high Galactic latitudes. If the
intrinsic population is cosmological this might indicate the Galaxy is
obscuring FRBs near the plane, but it is not clear what the cause
might be\cite{psj+14}. We can perform a check, in analogy to ones
performed in the early days of $\gamma$-ray burst studies: by
determining dipolar and quadrupolar moments. The statistics $\langle
\cos\theta \rangle$ and $\langle \sin^2b-1/3 \rangle$ should be
Gaussian distributions centered at zero, for an isotropic
distribution, where $\theta$ is the angle between the Galactic centre
and an FRB, and $b$ is the Galactic latitude of the FRB. It is
straight-forward to show that with as few as $10$ FRBs these
statistics behave in their Gaussian limits, and as such their standard
deviations are given\cite{bri93} by:
$\sigma_{\langle\cos\theta\rangle} = (1/3N_{\mathrm{FRB}})^{1/2}$ and
$\sigma_{\langle\sin^2b - 1/3\rangle} =
(4/45N_{\mathrm{FRB}})^{1/2}$. From this simple test, it seems that
the distribution of FRBs from Parkes is isotropic, but this quick
comparison does not take a number of important completeness factors
into consideration. For example this distribution must also be
weighted by the total time spent on sky across all
latitudes. Historically, a disproportionately large amount of time was
spent searching the plane of the Galaxy, where the rate of pulsar
discovery are highest. While some FRBs have been found at low Galactic
latitudes\cite{kkl+11,sch+14}, searches in other surveys specifically
covering the Galactic Plane region have been unsuccessful at finding
new bursts\cite{psj+14}.

The combination of number of bursts and total time on sky will ultimately 
be the only reliable way of interpreting the total FRB rate and determining
the degree of isotropy of the detected bursts. However, given the small sample 
currently available even a proper consideration of the corrections above
prevents one from drawing a firm conclusion. A significant increase in the 
total population will be essential.

Moreover, detector-dependent selection effects also play an important
role in the detectability of FRBs, not only at different telescopes
but also for bursts with different widths and total dispersion
measures. Recent works have shown\cite{kp15} that the algorithms used
in some single pulse search codes can cause FRBs in the data to be
missed, or their signal-to-noise ratio to fall below the threshold for
a candidate event to be positively flagged by the search codes. In
combination with other selection effects -- such as the preferential
sensitivity of many searches to pulses with narrower widths\cite{kp15}
and a suspected reporting bias in the Galactic plane\cite{frbrrat16}
-- that means that many surveys are not equally sensitive to all kinds
of FRB pulses. A large enough sample of FRBs observed from different
telescopes and backends is needed in order to constrain the severity
of these effects.

\section{The Future}

At present the discovery rate at Parkes, with the typically available
time-on-sky, is $\sim 5$ FRBs per year. This rate may be much higher
if more dedicated search time is available, such as the planned use of
the telescope for the {\it Breakthrough Listen} initiative, which will
be sensitive to dispersed radio pulses. During the lifetime of this
project, over the next $\sim 5$ years, it is in principle possible to
perform FRB searches commensally with pulsar timing and SETI searches
so as to increase the time on sky. Longer term plans to put a PAF at
the focus of the Parkes telescope may play into the selection effects
discussed above, as using the current generation of PAFs necessarily
sacrifices sensitivity for field of view. While increased
field-of-view will certainly be beneficial, any reduction in
sensitivity may mean only finding the brightest (and/or lowest
redsfhift) of the eventss, perhaps hindering understanding of the
population in its entirety\cite{caleb16}.

The current population of FRBs already offers a rich and exciting
field of study. The 16 published events to-date have a variety of
properties and features\cite{frbcat} and future discoveries hold great
promise. As new instruments come on-line to search for FRBs in the
coming years the population is expected to grow rapidly. Robust
population studies should be possible within the next five
years\cite{caleb16} and many aspects of the FRB mystery may be solved
in that time. Parkes remains a premiere instrument for FRB
searching. In fact, advances in survey strategy and data processing
have secured the SUPERB survey a leading role in this field in the
near future. In particular, almost real-time detections in the SUPERB
survey will enable us to answer the critical questions as to the
origins of FRBs, their distribution throughout the Universe, and their
usefulness as physical probes.

Ultimately, the future of FRB searches lies with the Square Kilometer
Array (SKA).  Searches with SKA1-MID are expected to yield several
FRBs per week\cite{mkg+15} and, much like the \textit{Swift} telescope
did for GRBs, bring us in to an era of regular real-time
discovery. FRBs have never been detected below 700 MHz, but if indeed
they are visible at lower frequencies SKA1-LOW will also be a powerful
tool for providing discoveries and for understanding the FRB
population.

\section*{Acknowledgments}
The author would like to thank the SUPERB collaboration for the work
they have put in during 2014 and 2015, and in particular to Ewan Barr
and Andrew Jameson. The author would like to additionally thank Andrea
Possenti and Emily Petroff for their help with the preparation of
these proceedings.

\end{document}